\documentclass{ws-ijmpa}

\begin{document}
\def\draftnote{}

\markboth{M. H. Reno}
{Deep Inelastic Neutrino Interactions}

%
%

\title{DEEP INELASTIC NEUTRINO INTERACTIONS
}

\author{\footnotesize MARY HALL RENO\thanks{To appear in the Proceedings of the 8th Workshop
on Nonperturbative Quantum Chromodynamics, 7-11 June 2004, Paris, France.}}

\address{Department of Physics and Astronomy,
University of Iowa\\
Iowa City, Iowa 52242 USA
}

\maketitle


\begin{abstract}
Next-to-leading order QCD corrections, order 1 GeV mass corrections
and the role of a strangeness asymmetry and isospin violation in the
$x$ dependence of parton distributions are evaluated in the
context of the neutrino-nucleon cross section. Their contributions
to evaluations of the weak mixing angle using the Paschos-Wolfenstein
relation are discussed.

\end{abstract}

\section{Introduction}    

Neutrino scattering with nucleons is a well-studied phenomenon in
a variety of energy regimes.\cite{pdg} Because of the results
on neutrino oscillations determined by the Super-Kamiokande experiment
and others from their studies of atmospheric neutrino fluxes,\cite{gg} tau neutrino
scattering as well as muon neutrino and electron neutrino scattering
is considered. In another arena of neutrino physics, the
NuTeV experimental collaboration's precision work on neutrino
scattering
and their determination of $\sin^2\theta_W$ has led to a
reexamination
of the neutrino cross section. Their result,\cite{nutev} that
\begin{equation}
\sin^2\theta_W = 0.2277\pm 0.0013\pm 0.0009\ ,
\end{equation}
is in contrast to the world
average without NuTeV:\cite{global} $0.2227\pm 0.0004$, a difference
of $\sim 0.005$.
In both of these cases, one is led to reconsider
the role of
next-to-leading order (NLO) perturbative QCD and  the role of the mass corrections
of order $O$(GeV). Here, we focus on the case of the NuTeV result for
$\sin^2\theta_W$, discussing the impact of the perturbative QCD
corrections\cite{kr,de,mm,davidson}
and mass corrections,\cite{kr} and considering the role of
a strangeness asymmetry and isospin violation in
the $x$ dependence of the parton distribution
functions in resolving the discrepancy.\cite{mm,davidson,krt}  

\section{NLO QCD and Mass Corrections}

The starting point for the neutrino-nucleon cross section,
including NLO QCD and mass corrections, is the differential
cross section in terms of $x=Q^2/(2P\cdot q)$,
$y=(E_\nu-E_\ell)/E_\nu$ and $q=p_\nu-p_\ell$ with $q^2=-Q^2$.
For $M$ representing the nucleon mass and $m$ the lepton mass, the
charged current differential cross section is
\begin{eqnarray}
\frac{d^2\sigma_{CC}^\nu(\nu N)}{dx\, dy} & = & 
\frac{G_F^2 M E_\nu}{\pi (1+Q^2/M_W^2)^2}
\Biggl[\Biggl(xy^2+\frac{m^2 y}{2ME_\nu}\Biggr) F_1^{TMC}\nonumber \\
&+& \Biggl( 1-\frac{m^2}{4E^2}-y-\frac{Mxy}{2E_\nu}\Biggr) F_2^{TMC}
+\Biggl(x-\frac{xy^2}{2}-\frac{m^2 y}{4ME_\nu}\Biggr) F_3^{TMC}\nonumber\\
&+& \Biggl( \frac{m^2y}{2ME_\nu}+\frac{m^4}{4 M^2 E_\nu^2 x}\Biggr)F_4^{TMC}
-\frac{m^2}{ME_\nu} F_5^{TMC}\Biggr] \ .
\end{eqnarray}
Lepton and target masses also enter through the limits on $x$ and $y$.
Charm mass corrections from $W^+ + s\rightarrow c$ are 
incorporated in the structure functions.\cite{kr,gott,gkr}
QCD and target mass corrections are included in the structure functions
$F_i^{TMC}=F_i^{TMC}(\xi,Q^2)$, where $\xi$ is the Nachtmann variable
defined by 
\begin{equation}
{1\over \xi}={1\over 2x}+\sqrt{{1\over 4x^2}+{M^2\over Q^2}}\ .
\end{equation}
The target mass corrected structure functions have target mass
dependence through $\xi$,  and they 
have corrections related to the mismatch between
partonic and hadronic tensor expansions of the interaction vertex squared.\cite{aot}
The final 
$M^2/Q^2$ corrections come from parton intrinsic
transverse momentum
effects which are limited by $M$.\cite{efp} Combined, these give
the same results as the operator product expansion [OPE] discussed
by Georgi and Politzer.\cite{gp}
The results, for example, for $F_2^{TMC}$ is
\begin{eqnarray}
F_2 ^{TMC} &=& 2\, \frac{x^2}{\rho^3}{{\cal F}_2(\xi ,Q^2)\over \xi}+ 
12\,\frac{M^2}{Q^2}\,\frac{x^3}{\rho^4}
\int_\xi^1 d\xi '{ {\cal F}_2(\xi ',Q^2)\over \xi'}\\ \nonumber
& & \quad\quad + 
24\, \frac{M^4}{Q^4}\, \frac{x^4}{\rho^5} 
\int_\xi^1 d\xi ' \int_{\xi '}^1 { {\cal F}_2(\xi '',Q^2)\over
\xi ''} \ ,
\end{eqnarray}
in terms of $\rho^2=1+4M^2 x^2/Q^2$ and ${\cal F}_2=q(\xi,Q^2)+\bar{q}(\xi,Q^2)$.

\section{Approach to $\sin^2\theta_W$}

The NuTeV experimental approach to extracting $\sin^2\theta_W$ uses
both $\nu_\mu N$ and $\bar{\nu}_\mu N$ scattering.
Their analysis uses
correlated  the correlated $R_{\nu}$ and  $R_{\bar{\nu}}$ 
measurements where 
\begin{equation}
R_{\nu,\bar{\nu}} \equiv
\frac{\sigma_{NC}^{\nu ,\bar{\nu}}}{\sigma_{CC}^{\nu,\bar{\nu}}}\ 
\end{equation}
in terms of the charged current (CC) and neutral current (NC) cross
sections.
A particularly useful theoretical quantity is 
\begin{equation}
\quad R^-\equiv \frac{\sigma_{NC}^\nu -\sigma_{NC}^{\bar{\nu}}}
{\sigma_{CC}^\nu -\sigma_{CC}^{\bar{\nu}}} \,
\end{equation}
where many uncertainties cancel. In addition, $R^-$ is
fairly independent of energy: the beam of neutrinos (and
antineutrinos) is distributed in energy. 
Theoretically, with several approximations:
\begin{itemize}
\item
assume isoscalar nucleons with
$u(x)=d(x)\equiv q(x)$ and the usual isospin relations
between up and down quark distributions in the proton and
neutron,
\item
neglect the
charm mass  so $s\rightarrow c$ not
suppressed (Cabbibo angles not relevant)
\item
neglect target masses  and work in leading order QCD
\item
take $s(x)=\bar{s}(x)$, etc., for the sea quark distributions,
\end{itemize}
one gets the Paschos-Wolfenstein relation\cite{pw}
\begin{equation} 
\label{eq:rminus}
R^-=\frac{(1-2\sin^2\theta_W)}{2}\ .
\end{equation}
With these approximations, Eq. (\ref{eq:rminus}) is
independent of the limits of integration on $x$ and $y$ which could
be used to mock up experimental cuts.

In terms of an actual
measurement of $R^-$, 
pure neutral current and charge current event samples are not possible.
In addition, there are the effects of cuts. The electron neutrino background
is subtracted. While the NuTeV measurement is not a {direct}
measurement of $R^-$, it is a useful theoretical effort to examine NLO QCD
corrections to $R^-$ to assess their impact on the extraction of
$\sin^2\theta_W$.

\section{Corrections to $R^-$}

\subsection{NLO QCD}
We first discuss the NLO perturbative QCD corrections to
$R^-$ including target
mass,
lepton mass and charm mass
corrections. 
As noted above, it is only with some approximations that 
$R^-$ is simply related to $\sin^2\theta_W$ as in Eq. (\ref{eq:rminus}),
so our approach is to look at the full NLO QCD corrections to 
$1/2-R^-$ and to compare with the LO evaluation. An approximate
moment analysis of the NLO corrections appears
in Refs. 6-8.  The full
evaluation of the NLO corrections leads to the results in Table 1. We show
results with an input $\sin^2\theta_W$ of 0.2227, using the
Gluck, Reya and Vogt parton distribution
functions\cite{grv} (PDFs), and the CTEQ6 PDFs,\cite{cteq6} including the
40 sets with individual variations in the 20 parameters
in the sets to estimate the error. To account for the
fluxes of neutrinos and antineutrinos, 
\begin{equation}
\label{eq:flux}
\sigma_{\rm NC, CC}^{\nu, {\bar \nu}}
= \frac{
\int d E_{\nu , {\bar \nu}}\ d \sigma_{\rm NC, CC}^{\nu, {\bar \nu}}
\ \Phi (E_{\nu {\bar \nu}})\ }{
\int d E_{\nu , {\bar \nu}}\ \Phi (E_{\nu {\bar \nu}})
}
\end{equation}
for incident neutrino/antineutrino flux $\Phi$. Details of the flux
used appear in Ref. 5.
To mimic some experimental cuts, we take 
${ 20\ {\rm GeV} < y E_{\nu , {\bar \nu}} < 180\ {\rm GeV}}$
in Eq. (\ref{eq:flux}). Table 1 shows that the NLO corrections
cannot account for the discrepancy between the NuTeV evaluation of
$\sin^2\theta_W$ and the world average.

\begin{table}[ht]
\tbl{NLO perturbative QCD corrections to $R^-$.$^5$}
{\begin{tabular}{@{}ccc@{}} \toprule
Input PDF & Input $\sin^2\theta_W $  & $\frac{1}{2}-R^-$\\
\colrule
GRV LO & 0.2227 & 0.2192 \\
GRV NLO & 0.2227 & 0.2192 \\
CTEQ6 NLO & 0.2227 & 0.2196 \\
 &  & $\pm 0.0005$ \\ \botrule
\end{tabular}}
\end{table}

\subsection{Strange sea asymmetry}

So far, we have assumed isospin symmetry and $q(x)=\bar{q}(x)$, however,
one can relax the second condition for strange sea. It is reasonable
that
$s(x)\neq \bar{s}(x)$ since the $s$ can arrange itself in
mesons and baryons, while the $\bar{s}$ can only go into meson fluctuations.\cite{ssbar}
The condition that the net strangeness of the nucleon vanishes must still be
satisfied:
\begin{equation}
\int _0^1 dx (s(x)-\bar{s}(x)) = 0 \ ,
\end{equation}
However, one can have
\begin{equation}
\left[ S^-\right] \equiv \int dx\ x\ \left[s(x) - {\bar s}(x)\right] \neq 0 \ .
\end{equation}
In terms of $[S^-]$, the implication for $R^-$ is that 
\begin{equation}
R^- \simeq {1\over 2} -\sin^2\theta_{\rm W} -\Bigl( {1\over 2} -{7\over 6}\sin^2
\theta_{\rm W}\Bigr) {[S^-]\over [Q^-]}
\end{equation}
with isoscalar up or down quark distributions $q(x)$ contributing via
$[Q^-]=\int x[q(x)-{\bar{q}}(x)] dx$.
A positive value for $[S^-]$ works in the direction to moderate the
disagreement between NuTeV and the world average values of $\sin^2\theta_W$.

Olness et al.\cite{olness} have performed a global analysis
of dimuon production
[$W^+ s \rightarrow c$
and  $W^- {\bar s} \rightarrow {\bar c}$ with semileptonic charm
decay] together with the other PDFs. Their results 
favor\cite{barone} $[s(x) - {\bar s}(x)]<0$ at 
low-$x$, and $[s(x) - {\bar s}(x)]>0$ at large-$x$, so
$[S^-]>0$. 
The allowed values for $[S^-]$ range between\cite{olness}
\begin{equation}
\label{eq:sminus}
-0.001<\left[ S^-\right]<0.004.
\end{equation}
A recent paper by Catani, de Florian, Rodrigo and
Vogelsang\cite{catani} shows that an asymmetry of $[S^-]\sim -5\times
10^{-4}$ is generated by NNLO perturbative evolution, even with a LO
value of $[S^-]=0$. The implication is that the strangeness asymmetry
must be of nonperturbative origin if it is as large as in
Eq. (\ref{eq:sminus}). We note that the NuTeV experimental collaboration reports
a small negative value for $[S^-]$.\cite{smalls}

Using our full NLO QCD evaluation of the cross sections, we
evaluate $R^-$ using several of the PDFs with a strangeness asymmetry
from Ref. 19 (labeled
A, B, C, B$^+$ and B$^-$) and compare them to the central value  of
$R^-$ using  the
CTEQ6 PDFs.
Defining
$$
\delta R^-\equiv  R^-_{\{{\rm A,B,C,B^+,B^-}\}} - R^-_{{\rm CTEQ6}} ,
$$
our results appear in Table 2. A positive value of $[S^-]$ gives
a negative value for $\delta R^-$.
Our conclusion is that a
nonperturbative input of $[S^-]>0$ consistent with
a global PDF analysis could explain the discrepancy in the 
$\sin^2\theta_W$ measurements.\cite{davidson,krt,mrst}

\begin{table}[t]
\tbl{Shifts in $R^-$ calculated$^9$ with the PDF sets of Olness et al.$^{19}$
compared to the CTEQ6M set with $[S^-]=0$.}
{\begin{tabular}{cccc}
\toprule
fit \quad &
\quad $\left[S^-\right] \times 100$ \quad &
\qquad $\delta R^-$ \qquad \\
\colrule
${\rm B}^+$& 0.540     & -0.0065\\ 
A        \hphantom{0}  & 0.312       & -0.0037 \\ 
B    \hphantom{0}      & 0.160  & -0.0019 \\ 
C   \hphantom{0}       & 0.103       & -0.0012 \\ 
${\rm B}^-$&-0.177 &  0.0023 \\ \botrule
\end{tabular}}
\end{table}

\subsection{Isospin violation}

Isospin symmetry violation,\cite{londergan} in which the valence
quark distributions do not obey the symmetry of
$u_V^n(x)\neq d_V^p(x)$, etc., could also account for the discrepancy
in $\sin^2\theta_W$. One finds approximately that
\begin{equation}
$$\delta R_I^- \simeq -\biggl(\frac{1}{2} -\frac{7}{6}\sin^2\theta_W
\biggr) \frac{[D_N^- -U_N^-]}{[Q^-]},\  \ [D_N^-]=\frac{1}{2}\bigl(
[D_p^-]+[D_n^-]\bigr) 
\end{equation}
The MRST\cite{mrst} PDFs yield
\begin{eqnarray}
\label{eq:isospin}
-0.007 < &\delta R_I^-& < 0.007\\ \nonumber
\delta R_I^- &=& -0.0022\quad{\rm for\ the\ best\ fit}\ ,
\end{eqnarray}
with the best fit again working in the direction to moderate
the discrepancy in $\sin^2\theta_W$ values.

\section{Summary}

NLO QCD can't account for the discrepancy between NuTeV and other
experimental evaluations of $\sin^2\theta_W$,
however,
a strange-antistrange asymmetry that preserves net strangeness zero in
the nucleon, consistent with a global PDF analysis of the data, moderates the discrepancy. 
Isospin violation may also come into play at the same level. Our results
suggest that the Weinberg angle measurements may be accommodated
within the standard model as long as parton distribution functions
show strangeness and/or isospin asymmetries of nonperturbative origin at the level
described in Table 2 and Eq. (\ref{eq:isospin}).

\section*{Acknowledgments}

This work is funded in part by the U.S. Department
of Energy contract  
DE-FG02-91ER40664. The author acknowledges S. Kretzer, W.-K. Tung,
J. Pumplin, F. Olness and D. Stump for their contributions to the work
presented here.

\end{document}